# 金属修饰 SiO$_2$ 基底表面对吸附石墨烯电子特性的影响

沈磊，江帆，肖蒙、张瑞，余牧溪，缪灵\*，江建军

(华中科技大学 电子科学与技术系 430074)

本文采用第一性原理计算，研究了不同金属修饰的 SiO$_2$ 基底表面对吸附石墨烯的结构稳定性及电子能带结构的影响。计算了金属修饰基底表面的不同体系形成能和吸附在基底表面上石墨烯的能带结构。结果表明，SiO$_2$ 基底 O 面吸附金属原子比 Si 面更为稳定，且用金属修饰的 SiO$_2$ 基底比通常用 H 钝化表面的基底更容易吸附石墨烯。在 Co 等金属修饰表面的 SiO$_2$ 基底上，石墨烯的电子特性保持较好，而当 Fe 修饰时，石墨烯的电子特性被明显的破坏。磁性金属原子的修饰使石墨烯能带结构在费米面附近发生分裂，一条保持原有的特性，而另一条打开了 1-2 eV 的带隙，显示出了半金属性的特性。

**关键词**：石墨烯，纳米带，电子特性，密度泛函

## 1. 引言

石墨烯（Graphene）自 2004 年被发现以来，由于具有新奇的物理性质和广泛的应用前景，成为近年来凝聚态物理领域研究的热点，并有可能成为构造下一代纳米电子器件的基本材料[1-3]。纯净石墨烯是零带隙半导体[1]，具有诸如无质量的狄拉克费米子、室温量子霍尔效应、弹道输运等一系列的物理性质[3]，使其在器件方面有各种应用，如研制能够发展传统半导体器件的小型、低功耗和具有更高工作频率的石墨烯晶体管，并有可能应用于将来的纳米集成电路中[1-5]。

近几年来，人们已经在石墨烯的制备方面取得了积极的进展，发展了机械剥离、晶体外延生长、化学氧化、化学气相沉积和有机合成等多种制备方法[6-9]。但是，在石墨烯通往规模化的工业生产和科研应用的道路上，还面临着一个重要的问题，就是石墨烯所依附的基底的选取。石墨烯仅由单层碳原子组成，如果不依附在一定的基底上，很难进行规模化的生产和应用。不同类型的基底以及不同类型的结合，都会对原有孤立的石墨烯的原子结构及电子特性产生不同程度的影响，甚至会大幅度的改变原有性质[14]。

以 SiO$_2$ 作电介质的绝缘基底被广泛应用在在集成电路生产与制备中[24]。已有大量的关于石墨烯片层在 SiO$_2$ 基底上生长状况的实验和理论研究，揭示了石墨烯-基底系统的形态结构和许多优异性质[10-16]。Kang 等人[14]研究了石墨烯片层和 SiO$_2$ 基底表面之间

---

的相互作用，发现当石墨烯放置在 O 面时，石墨烯和基底表面形成强烈的 C-O 键，极大的改变了石墨烯的电子特性，打开了带隙；而当石墨烯放在 Si 面时，石墨烯的性质几乎不受到影响。Shemella 等人[13]研究发现当 $SiO_2$ 基底 O 面用 H 钝化后，放置在其上的石墨烯的性质得以保持。这些研究表明石墨烯和基底间界面形态对生长在基底上石墨烯的电子特性有很大的影响，而当基底表面经过化学修饰之后，放置在其上的石墨烯的电子特性就有可能被调制。

本文工作旨在通过第一性原理计算，系统地研究了不同金属修饰 $SiO_2$ 基底表面，包括 Ti、Ca、Ni、Mn、Co、Fe、Cr、K 等金属，分别考虑了两种 O 面和一种 Si 面，以探究不同金属的最稳定位置。通过对石墨烯放置在不同界面形态下形成能的对比，发现石墨烯-基底系统的最稳定位置结构并讨论了不同结构形态对石墨烯电子特性的影响。

## 2. 计算方法和结构模型

几何结构优化和电子结构的计算采用了基于密度泛函理论（DFT）[17,18]的 SIESTA 软件包[19]，它采用了原子轨道线性组合（LCAO）方法[20]。在进行结构驰豫和电子结构的计算中，采用局域密度近似 LDA（local-density-functional approximation）泛函形式[21]处理交换相关势能,计算使用完全非局域形式(Kleinman-Bylander)的标准守恒赝势[22]。通过 Monkhorst–Pack 方法[23]在简约布里渊区中产生 8×8×1 个 k 点。收敛判据设置为每个原子受力小于 0.05 eV/ Å，赝势平面波截断能选取为 200 Ry，以达到计算效率和精度的平衡。

本文选取α-石英 $SiO_2$ 晶体，这是一种较为常见的 $SiO_2$ 晶体。取其（0001）面作为基底模型，底部用 H 饱和。为了防止 z 方向的周期影响，模型上方建立了 20Å 的真空层。由于晶格参数的差异，基底与石墨烯之间存在晶格匹配的问题，$SiO_2$ 晶格参数 a=b=4.913Å，与 2×2 的石墨烯（a=b=4.92Å）相差仅为 0.1%，计算确认了匹配后的石墨烯性质几乎没有受到影响。

本文计算中选取了 $SiO_2$（0001）面的三种表面位置（两种 O 面和一种 Si 面，如图 1），在其上方分别吸附金属原子。本文主要分析了三种石墨烯片层放置的位置，分别是金属原子位于石墨烯晶格六边形正中央下方（H）、C-C 键正下方（B）和 C 原子正下方（T）（如图 2）。

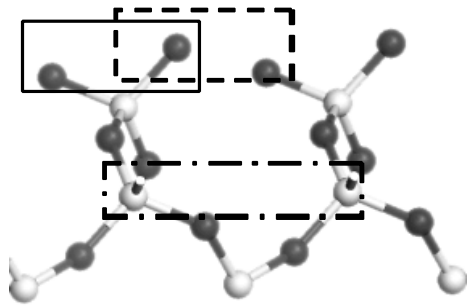 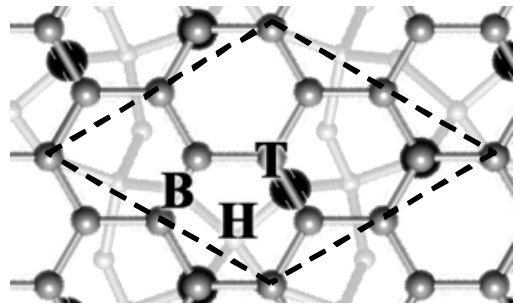

图 1 三种表面位置（实线框为 O1 表面，虚线框为 O2 表面，点划线为 Si 表面），金属原子位于两个表面原子上方　　图 2 石墨烯与 SiO2 基底晶格匹配（灰色为 C 原子，黑色为表面 O 原子，虚线框所指示的为调整后的 2×2 石墨烯晶胞）

## 3. 计算结果与讨论

### 3.1. SiO₂ 基底吸附金属模型的稳定性

本文计算采用 SiO₂ 基底的三种表面吸附不同金属原子，包括碱金属中的 K、Ca，过渡金属中的 Ti、Cr、Mn、Fe、Co、Ni、Cu，其中包括磁性元素 Mn、Fe、Co、Ni。为了在三种表面上为每种金属寻找一个较为稳定的位置，对于每一种表面选取对称性较好的位置进行结构优化并计算形成能，判断稳定性。采用如下的形成能定义：

$$\triangle E = E_{sm} - E_s - E_m$$

其中 $E_{sm}$ 表示基底吸附金属体系的总能，$E_s$ 表示孤立的基底总能，$E_m$ 表示金属原子的总能。对 Ti 等金属进行计算，得到其三种表面位置的形成能，见表 1。

表 1 SiO2 表面不同位置吸附金属原子形成能比较(eV)

| 金属原子 \ 表面位置 | O1 | O2 | Si |
| --- | --- | --- | --- |
| Ti | -5.1900 | -14.0460 | -12.5100 |
| Ca | -5.1178 | ---- | -0.7261 |
| Ni | -5.0133 | -11.5215 | -9.3535 |
| Mn | -11.3419 | -13.1269 | -5.1120 |
| Cr | -16.3191 | -18.5162 | -10.9645 |
| Fe | ---- | -12.7110 | -5.3870 |
| Co | ---- | -12.1480 | -5.2710 |
| K | -3.848 | -5.439 | -0.528 |

根据表中的形成能可以看出，Ca 原子吸附在 O1 表面较为稳定，而 Ti、Ni、Cr 等原子吸附在 O2 表面较为稳定。对于以上计算的大多数金属而言，金属原子吸附在 SiO2

基底的 O 表面比吸附在 Si 表面更加稳定，这是因为金属原子吸附在 O 表面能和表面 O 原子产生强烈相互作用，形成较强的金属-氧化学键；而当金属原子吸附在 Si 表面时，由于金属原子和基底的表面 Si 之间的相互作用比较弱，使得形成的基底吸附金属体系不如 O 表面基底稳定。对大多数金属，吸附在 O2 表面的体系比吸附在 O1 表面的体系要更稳定一些。

### 3.2. 金属修饰 SiO$_2$ 基底表面吸附石墨烯的稳定性

得到金属吸附的稳定位置后，进一步研究石墨烯放置在该表面的稳定性。在稳定吸附各金属原子的 SiO$_2$ 表面放置石墨烯片层，使金属原子分别处于石墨烯结构中的 H、B、T 位置（如图 1 所示）。初始石墨烯片层和金属原子的 z 方向高度差均设置为 2 Å。

表 2 石墨烯在金属修饰基底表面的形成能(eV)

| 吸附原子 \ 石墨烯位置 | H | B | T |
|---|---|---|---|
| Ti（O2） | -3.4349 | -3.1273 | -2.9834 |
| Ca（O1） | -1.4664 | -1.2651 | -1.2430 |
| Ni（O2） | -2.1565 | -2.0172 | -1.6684 |
| Mn（O2） | -8.6918 | -8.7524 | -8.9225 |
| Cr（O2） | -0.9959 | -1.0900 | -1.0153 |
| Fe（O2） | -3.5319 | -2.6247 | -2.6469 |
| Co（O2） | -4.1284 | -4.2524 | -4.2134 |
| H | | -0.6318 | |

在以上考虑的三种位置中，有着最大吸附能的位置被称为最稳定位置。由图表 2 可知，大多数金属最稳定位置在 H(石墨烯 C 六边形的正下方)，这是由于 H 位置的金属原子与石墨烯六圆环的六个 C 等距，六个 C 原子分散了与金属的相互作用力，亦即较稳定。另外，Mn 最稳定位置可能在 T（即石墨烯中 C 原子正下方），Cr 最稳定位置可能在 B（即石墨烯 C-C 键中央正下方）。对于 B、H、T 三种位置，优化后的形成能相差不大，在一些情况下，这种差距可以忽略。

与用 H 饱和 SiO$_2$ 表面放置石墨烯片层的形成能相比较，可以看出形成能绝对值较低的 Cr 原子，其数值也为 H 饱和系统的形成能绝对值的两倍左右。由此可知，金属修饰后的 SiO2 基底可以和石墨烯片层产生更强的结合，结构更加稳定。

同时，与研究独立的金属修饰石墨烯相比，本文所研究的系统具有更高的稳定性和可行性。

## 3.3. 金属修饰 SiO$_2$ 基底对石墨烯性质和能带的影响

本文通过计算，得出各种金属原子吸附在稳定 SiO$_2$ 表面与石墨烯系统的能带结构。从能带结构可以观察出，金属修饰基底对石墨烯能带结构的影响有两种典型情况。

一种是石墨烯的能带结构被良好的保持，导带底与价带顶交于一点且位于费米能级附近（图3），典型金属如 Co。此外，K、Ca、Mn、Cr 等金属原子也较好的保持了石墨烯的原始能带结构。

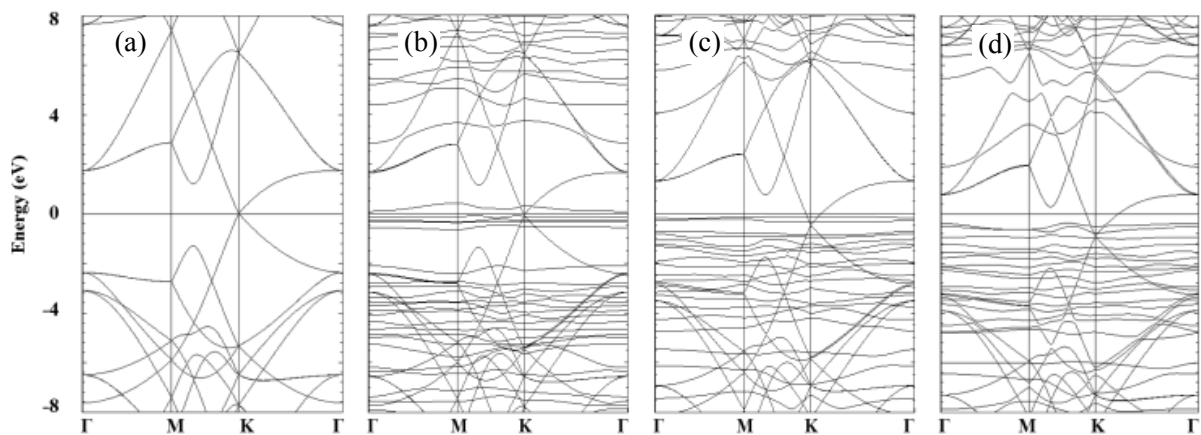

图 3 (a)石墨烯能带图 (b)Co、(c)K、(d)Ca 修饰 SiO2 基底放置石墨烯片层能带图

为了研究费米能级附近出现的许多平带，画出 Co 原子 3d 轨道的 PDOS 图（图 4(b)）。由图中可以看出，费米能级附近的平带为吸附金属原子引入，并非石墨烯片层的能带。

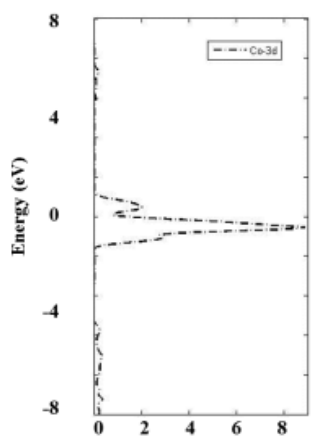 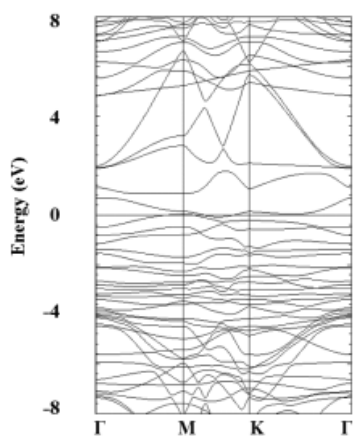

图 4 Co 吸附模型 Co 原子 3d 轨道 PDOS 图　　　图 5 Fe 原子吸附 SiO2 基底放置石墨烯能带图

从图 3 中可以看出，所研究的多数金属修饰基底上石墨烯能带结构被保持，但费米能级被不同的程度的移动，大部分被向上移动。为了验证石墨烯和基底之间是否发生了

电子转移,表 4 中列出了 Co、Ca 金属吸附系统中金属原子核 C 原子的布居数。

表 4 石墨烯在金属修饰基底表面的原子布居数

| 吸附原子 | 金属原子 | C1 | C2 | C3 | C4 | C5 | C6 |
|---|---|---|---|---|---|---|---|
| Co(O2) | 8.453 | 4.017 | 4.014 | 4.019 | 4.010 | 4.011 | 4.017 |
| Ca(O1) | 0.968 | 4.070 | 4.059 | 4.051 | 4.054 | 4.066 | 4.051 |

由表中数据可以看出,放在金属修饰基底体系表面的石墨烯中 C 原子的布居数和独立的石墨烯中 C 原子布居数相差不大,这说明石墨烯和基底之间并没有发生明显的电荷转移,使得石墨烯的性质能够较好的保持。这一结果和图 5.1(b)(c)(d)中能带结构中费米能级附近石墨烯性质保持较好的现象是一致的。

另一种情况是石墨烯的能带结构被明显的破坏,典型金属如 Fe 等(见图 5)。由图可以看出,石墨烯能带结构上半部分可以看到,但下半部分几乎已经杂乱消失。

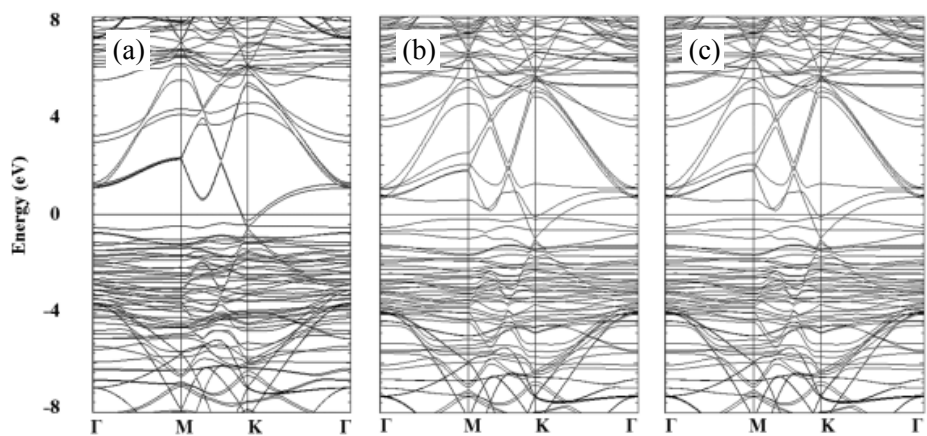

图 6 Ni、Mn、Cr 原子吸附 SiO2 基底放置石墨烯能带图(加自旋)

对于部分过渡金属(如 Ni、Mn、Cr),计算时加入自旋,计算结果能带分离为自旋向上和自旋向下两条能带(图 6)。从图中可以看出,石墨烯能带结构在费米面附近原有的相交能带分裂成两条,一条近似的保持了原有的特性,而另一条则有一个 1-2 eV 的带隙,产生自旋极化,显示出了半金属的特性。

## 4. 结论

采用基于密度泛函理论(DFT)的第一性原理计算,系统地研究了不同金属修饰 $SiO_2$ 基底表面,包括 Ti、Ca、Ni、Mn、Co、Fe、Cr、K 等金属,并分别考虑了两种 O 面和一种 Si 面,以探究不同金属的最稳定位置。结果表明 $SiO_2$ 基底 O 面吸附金属原子比 Si

面更为稳定。当石墨烯放置在不同金属的最稳定界面形态时，计算并比较了不同体系的形成能，结果表明用金属修饰的 $SiO_2$ 基底比通常用的 H 钝化表面的基底更易吸附石墨烯，形成的体系更加稳定。进而讨论了不同结构形态对石墨烯电子特性的影响，发现在 Co、K、Ca 金属修饰表面的 $SiO_2$ 基底上，石墨烯的电子能带结构得到良好的保持，而在 Fe 修饰时，吸附在基底上的石墨烯的电子特性被明显的破坏。当用磁性金属原子修饰基底表面时，考虑了自旋的影响，石墨烯能带结构在费米面附近发生分裂，一条保持原有的特性，而另一条打开了 1-2 eV 的带隙，显示出了半金属性。

## 参考文献